\documentclass[superscriptaddress,aps,prl,twocolumn]{revtex4-2}

\usepackage{bm}
\usepackage{dcolumn}
\usepackage{graphicx}
\usepackage[colorlinks=true, linkcolor=blue, citecolor=blue, urlcolor=blue]{hyperref}
\usepackage{mathrsfs}
\usepackage{multirow}
\usepackage{rotating}
\usepackage{url}
\usepackage{color}
\usepackage{amsmath}
\usepackage{amssymb}
\newcommand{\etal}{{\em et al.}}
\newcommand{\ie}{\textit{i}.\textit{e}.}
 
\begin{document}


\title{Fragile Unconventional Magnetism in RuO$_2$ by Proximity to Landau-Pomeranchuk  Instability}

\author{Zhuang Qian}
\thanks{These authors contributed equally to this work.}
\affiliation{Institute for Theoretical Sciences, Westlake Institute for Advanced Study,Westlake University, Hangzhou 310024, Zhejiang, China}
\author{Yudi Yang}
\thanks{These authors contributed equally to this work.}
\affiliation{Institute of Natural Sciences, Westlake Institute for Advanced Study,Hangzhou 310024, Zhejiang, China}
\affiliation{Key Laboratory for Quantum Materials of Zhejiang Province, School of Science, Westlake University, Hangzhou 310024, Zhejiang, China}
\author{Shi Liu}
\email{liushi@westlake.edu.cn}
\affiliation{Institute of Natural Sciences, Westlake Institute for Advanced Study,Hangzhou 310024, Zhejiang, China}
\affiliation{Key Laboratory for Quantum Materials of Zhejiang Province, School of Science, Westlake University, Hangzhou 310024, Zhejiang, China}
\author{Congjun Wu}
\email{wucongjun@westlake.edu.cn}
\affiliation{Institute for Theoretical Sciences, Westlake Institute for Advanced Study,Westlake University, Hangzhou 310024, Zhejiang, China}
\affiliation{Institute of Natural Sciences, Westlake Institute for Advanced Study,Hangzhou 310024, Zhejiang, China}
\affiliation{Key Laboratory for Quantum Materials of Zhejiang Province, School of Science, Westlake University, Hangzhou 310024, Zhejiang, China}
\affiliation{New Cornerstone Science Laboratory, Department of Physics, School of Science, Westlake University, Hangzhou 310024, Zhejiang, China}
\date{\today}

\begin{abstract}
Altermagnetism has attracted considerable attention for its remarkable combination of spin-polarized band structures and zero net magnetization, making it a promising candidate for spintronics applications. We demonstrate that this magnetic phase represents a case of ``unconventional magnetism," first proposed nearly two decades ago by one of the present authors as part of a broader framework for understanding Landau-Pomeranchuk instabilities in the 
spin channel, driven by many-body interactions.
By systematically analyzing the altermagnetism in RuO$_2$ with first-principles calculations, we reconcile conflicting experimental and theoretical reports by attributing it to RuO$_2$'s proximity to a quantum phase transition.  
We emphasize the critical role of tuning parameters, such as 
the Hubbard $U$,
hole doping, and epitaxial strain, in modulating quasiparticle interactions near the Fermi surface. This work provides fresh insights into the origin and tunability of altermagnetism in RuO$_2$, highlighting its potential as a platform for investigating quantum phase transitions and the broader realm of unconventional magnetism.
\end{abstract}

\pacs{
}
\maketitle

\newpage
The emerging research direction of altermagnetism ~\cite{Smejkal22p040501, Smejkal22p031042}, which is considered by many as a new type of itinerant magnetism~\cite{Igor22p040002}, has sparked considerable interests due to its unique properties and potential application in spintronics and magnetic memory devices~\cite{Amin24p348}. 
Unlike conventional antiferromagnetism (AFM), 
altermagnetism represents distinct features: 
It exhibits zero net magnetization arising from the coupling of sublattices with opposite spins 
related by lattice rotations 
rather than translation or inversion. 
Remarkably, altermagnetism combines characteristics of both ferromagnetism and AFM. 
It breaks the Kramers degeneracy, leading to spin-polarized band structures, while maintaining a zero magnetization. This duality enables various intriguing phenomena, including spin-polarized currents~\cite{Rafael21p127701}, anomalous Hall effects~\cite{Smejkal20peaaz8809, Feng22p735, Tschirner23p101103}, and potential applications in spin-transfer torque~\cite{Karube22p137201, Bai22p197202, Bose22p267, Bai23p216701}, all without the drawbacks of stray magnetic fields. Recent theoretical advancements, particularly the development of spin space groups integrating both spatial and spin operations~\cite{Xiao24p031037, Chen24p031038, Jiang24p031039}, have provided a comprehensive classification of these systems, spurring an intensive search for new altermagnetic materials.

\begin{figure*}
\centering    
\includegraphics[width=0.8\textwidth]{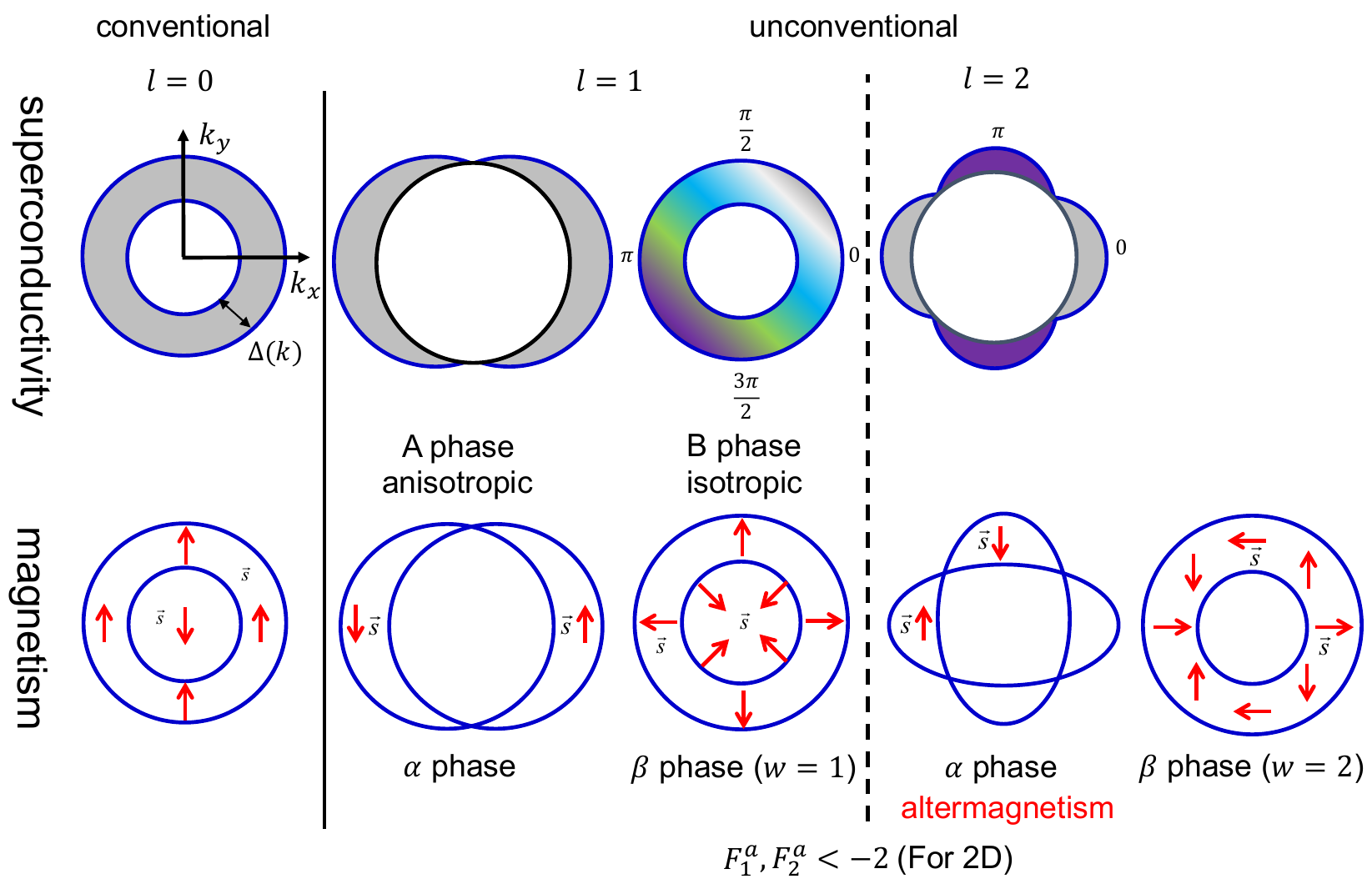}
    \caption{Comparison of superconductivity and magnetism. Upper panel: Conventional $s$-wave superconductivity, characterized by angular momentum $l=0$, and unconventional superconductivity with $l=1$ ($p$-wave) and $l=2$ ($d$-wave), where the superconducting gap function changes sign around the Fermi surface. Lower panel: Conventional ferromagnetic state, alongside unconventional $p$-wave and $d$-wave magnetism. The vector $\vec{s}$ represents the electron spin and {\it{w}} is winding number. For $p$-wave symmetry, both superconductivity and magnetism can exhibit isotropic or anisotropic phases. Unconventional $d$-wave magnetism arising form the LP instabilities in the spin channel corresponds to the phenomenon of $d$-wave altermagnetism.
    The schematic for $l=2$ unconventional magnetism is a replot of Fig.~1 from [\href{https://journals.aps.org/prb/abstract/10.1103/PhysRevB.75.115103}{Phys.~Rev.~B \textbf{75}, 115103 (2007)}] and is identical to the now-iconic illustration of altermagnetism.
    }
    \label{FigUncon}
\end{figure*}

It is not that uncommon that early discoveries remain relatively unknown and even overlooked.
We highlight that over two decades ago, Wu and Zhang introduced a mechanism for generating what Wu later termed {\it unconventional magnetism}~\cite{wu07website}
due to their symmetry properties similar to unconventional superconductivity under rotations. 
They are characterized by nonrelativistic momentum-dependent spin splitting in the absence of net magnetization~\cite{Wu04p036403}.
Furthermore, Wu, Sun, Zhang and Fradkin systematically developed the theory for this class of exotic states \cite{Wu07p115103}, including their symmetry
structures similar to spin-group in later
literature \cite{Xiao24p031037, Chen24p031038, Jiang24p031039}, topological defects and collective excitations.
Various properties are proposed for experiment tests, including temperature-dependent spin splitting bands detectable by angular-resolved photo emission and Shubnikov–de Haas  oscillations, spin-current generation from charge-current with tunable relative transport directions, coupling between strain and magnetization in the $d$-wave unconventional magnetic state, and resonance modes in the inelastic neutron scatting spectra 
\cite{Wu07p115103}.


The unconventional magnetism developed above is 
driven by electron interactions, based on the Landau-Pomeranchuk (LP) type 
instabilities in the spin channel with 
higher partial-waves denoted by orbital angular momentum number $l\ge 1$.
It is applicable to strongly correlated and nonrelativistic systems, explaining the emergence of spin-momentum locked Fermi surface distortions as a consequence of quantum phase transitions
beyond the relativistic spin-orbit coupling.
In contrast, while the spin-space-group-based symmetry analysis provides a framework for understanding the single-particle band structures in altermagnetism, it does not elucidate the origin of magnetic orderings, instead treating it as a pre-existing condition. 
As will be discussed below, altermagnetism falls into the same symmetry class of the spin channel LP instabilities with even $l$'s,
\ie, 
they could be smoothly connected by adiabatic evolutions. 


We first briefly discuss the study of LP instabilities within the framework of the Landau-Fermi liquid theory \cite{POMERANCHUK1959}.
The Landau-Fermi theory describes the low-energy behavior of interacting fermions in terms of long-lived quasiparticles near the Fermi surface. 
The effective interactions between quasiparticles can be parameterized by Landau parameters, denoted as  $F_l^{s}$  (the density channel) and $F_l^{a}$ (the spin channel).
When a Landau parameter is negatively large exceeding
a critical value, it triggers an LP instability, leading to a distortion of the Fermi surface. 
The nature of the instability depends on $l$ and whether it occurs in the density or spin channel as explained below.  
In the density channel,
a negative $F_2^s<-2$ in two dimensions (2D) leads to the formation of a nematic Fermi fluid phase associated with elliptically distorted Fermi surfaces, as investigated by Oganesyan 
\etal ~\cite{Oganesyan01p195109}.
In the spin channel, LP instabilities exhibit even richer physics. 
For $F_0^a<-1$, the system undergoes a ferromagnetic transition known as Stoner instability.
As for the $p$-wave case,
Hirsch proposed the emergence of spin-split states, characterized by the opposite 
displacements of the spin-up and spin-down Fermi surfaces about the center of the Brillouin zone~\cite{Hirsch90p6820, Hirsch90p6828}.  
Interestingly, Oganesyan \etal ~ speculated that $F_2^a<-2$ in 2D induces a spontaneous distortion of two Fermi surfaces for up and down spins into two orthogonal ellipses~\cite{Oganesyan01p195109}, an early hint of altermagnetism.

The theoretical framework developed by Wu and Zhang was the first systematic approach to studying LP instabilities in the spin channel~\cite{Wu07p115103}.
For simplicity, let us take the 2D case as an example. 
The order parameters at $l=1$ can be viewed as spin-currents.
It can be extended to arbitrarily high orbital partial waves [$F_l^a(l>0)$] as spin-multipole moments in momentum space.
Ginzburg-Landau free energy functionals are constructed to analyze the ordered phase patterns.
Two distinct types of ordered phases were identified, termed the $\alpha$ and $\beta$-phases,  which are analogous to the superfluid $^3$He-A and B phases,
respectively,
as shown in Fig.~\ref{FigUncon}. 
In the $\alpha$-phases, the Fermi surfaces of spin-up and spin-down electrons exhibit opposite anisotropic distortions.
In contrast, the Fermi surfaces in the $\beta$-phases remain circular while spin
configurations winds around Fermi surfaces.
Importantly, the symmetry structure of the Fermi surface in the $\alpha$-phase for $l=2$ 
is identical to that in the $d$-wave altermagnetism. 
In addition, the symmetry structures of spin current induced by charge current in the $d$-wave phase was also derived in Sect.~X in Ref.~\cite{Wu07p115103}.
It is worth mentioning that 
distorted Fermi surfaces
in the $d$-wave state~\cite{Wu07p115103}
and the spin current configuration shown in Wu's presentation in 2007~ \cite{wu07website} are the same as the now-iconic illustration of altermagnetism ~\cite{Smejkal22p040501, Smejkal22p031042}.
Furthermore, the residual symmetry of unconventional symmetries is analyzed containing composed operations of non-equivalent orbital and spin rotations, \ie, spin-space group type operations 
~\cite{Xiao24p031037, Chen24p031038, Jiang24p031039}. Unconventional magnetism was also generalized to orbital band systems to explain the nematic metamagnetism observed in Sr$_3$Ru$_2$O$_7$~\cite{Lee2009} and ultra-cold fermionic atom systems with large magnetic dipolar interactions~\cite{Li2012}, based on the Landau interaction strength. 


We seek a unified framework for understanding both unconventional magnetism and  altermagnetism from the symmetry perspective.
The classification of altermagnetism by the symmetry of the spin-degenerate nodal surfaces, such as $d$-wave, $g$-wave, or $i$-wave~\cite{Smejkal22p040501, Smejkal22p031042}, corresponds directly to the $\alpha$-phase of unconventional magnetism arising from LP instabilities with even $l$'s.
They share the same symmetry structures including spatial inversion and rotational symmetries both in spin and orbital channels.
Since time-reversal (TR) symmetry is broken, if lattice structures are further taking into account which was neglected in Ref.~\cite{Wu07p115103}, the $\alpha$-phases with even $l$'s will also exhibit magnetic structures in real space within a unit cell.
In contrast, unconventional magnetic states with odd $l$'s maintains TR symmetry, hence it is not covered by altermagnetism since no real space magnetic structures will be formed. 
This symmetry perspective highlights the deep connection between the two phenomena, offering insights into their shared theoretical underpinnings.


Understanding altermagnetism as an ordered phase resulting from a quantum phase transition naturally implies that this effect can be switched on and off near the critical 
point by tuning specific external parameters.
We propose that the ongoing controversy surrounding the altermagnetism of RuO$_2$ can be resolved by recognizing that its fragile altermagnetic state arises from its proximity to an LP instability.
The current understanding of magnetism in RuO$_2$ is briefly summarized below, highlighting its evolving characterization and ongoing controversies. 

RuO$_2$ was classified as a nonmagnetic metal based on early measurements of magnetic susceptibility~\cite{Ryden70p6058} and electrical resistivity~\cite{Lin04p8035}. However, recent neutron diffraction experiments revealed AFM ordering, reporting magnetic moments of approximately 0.05$\mu_B$ per Ru atom from polarized neutron diffraction
~\cite{Berlijn17p077201}. 
Resonant X-ray scattering confirmed these findings, establishing collinear AFM order and a N\'eel temperature above 300 K~\cite{Zhu19p017202}.
Both the anomalous Hall effect~\cite{Feng22p735, Tschirner23p101103} and spin-polarized currents~\cite{Karube22p137201, Bai22p197202, Bose22p267, Bai23p216701} in RuO$_2$ have been reported in experiments. 
However, significant inconsistencies remain regarding the AFM ordering in RuO$_2$. 
The magnetic moment extracted from polarized
neutron differaction is too small \cite{Berlijn17p077201} to explain the observed anomalous Hall effect~\cite{Feng22p735}. 
Recent muon spin rotation/relaxation ($\mu$SR) experiments, combined with density functional theory (DFT) calculations, have placed stricter upper limits on the magnetic moments: 1.4$\times$10$^{-4}\mu_B$ per Ru atom in bulk and 7.5$\times$10$^{-4}\mu_B$ in epitaxial thin films~\cite{Hiraishi24p166702, Philipp24arxiv}. Recent spin- and angle-resolved photoemission spectroscopy experiments have also failed to detect the expected spin splitting~\cite{Liu24p176401}. 

Theoretical investigations based on DFT + $U$ calculations have shown that the magnetic state of RuO$_2$ depends strongly on $U$ applied to Ru $4d$ orbitals, which reflects a strong correlation effect
arising from Coulomb repulsion.
Achieving an AFM ground state typically requires $U$ value above $1.2$~eV~\cite{Berlijn17p077201}, which is generally considered unphysical for a metallic system. Recent work suggests Ru vacancies, common in this material, could stabilize magnetism with a more reasonable $U$ below 1 eV, though the vacancy concentration needed is quite high, 0.4 holes per Ru, which is equivalent to 10\% Ru vacancies~\cite{Smolyanyuk24p134424}. 


By employing DFT calculations
(see details, 
in End Matter),
we systematically investigate the magnetic states by applying varying magnitudes of Hubbard $U$ corrections to the Ru $4d$ states, combined with equibiaxial epitaxial strains and hole doping. We find that achieving convergence of the ground-state magnetism at certain strain states and doping conditions requires an unusually dense $k$-point grid of 20$\times$20$\times$20 and electronic self-consistency convergence threshold of 10$^{-12}$ Ry. Our DFT-based approach examines the sensitivity of the magnetic properties to electronic correlations, modulated by $U$, strain, and the screening effects of itinerant carriers. 

We propose that the conflicting experiment-theory results of RuO$_2$ can readily be interpreted as evidence of fragile magnetism arising from the proximity to an LP instability. Given the many-body nature of the Landau parameter $F_2^a$, its precise determination through single-particle DFT calculations is impractical. However, it is reasonable to propose that $F_2^a$ depends on the strength of the on-site Coulomb repulsion, characterized by the $U$ value, as well as other external tuning parameters. These include in-plane equibiaxial epitaxial strain ($\eta$), which modulates the hopping integral ($t$) 
and thereby influences $U/t$.
The proximity to the LP instability would manifest as a pronounced sensitivity of the magnetic properties to these parameters. To illustrate this, we map out the phase diagram as a function of $U$ and $\eta$. For a given $\eta$, the in-plane lattice parameters are fixed to the values specified by the strain condition, while the atomic coordinates and out-of-plane lattice parameter are allowed to relax.

\begin{figure}[ht]
    \centering
    \includegraphics[width=0.45\textwidth]{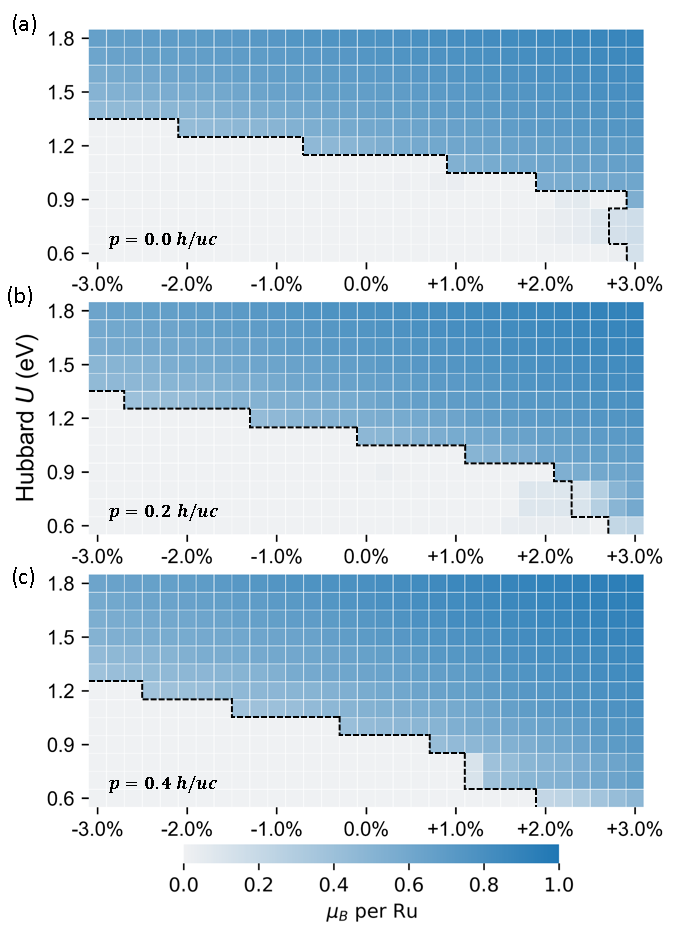}
    \caption{Heatmap of the magnetic order in RuO$_2$ as a function of the Hubbard parameter $U$ and equibiaxial epitaxial strain $\eta$, for hole doping levels of (a) $p = 0.0$, (b) $p = 0.2$, and (c) $p = 0.4$ holes per unit cell. The background color scales with the local magnetization on Ru atoms, with dashed lines indicating phase boundaries. Due to numerical convergence issues (see text) arising from proximity to the LP instability, only states with a local magnetic moment greater than 0.1 $\mu_B$ per Ru are considered AFM.}
    \label{PhaseBoundary}
\end{figure}

As shown in Fig.~\ref{PhaseBoundary}(a), the phase diagram reveals a transition (dashed line) between the non-magnetic state and the AFM state that features a local magnetic moment exceeding 0.1~$\mu_B$ per Ru. 
At zero strain, the critical 
value, $U^*$, is 1.2 eV, above which AFM ordering emerges. Under a tensile strain of 2\%, $U^*$ decreases to 1.0~eV. 
This behavior can be attributed to tensile strain reducing the hopping integral $t$, thereby allowing a smaller $U$ 
to drive the LP instability. A recently proposed Hubbard model on a square lattice with decorated next-nearest-neighbor hopping predicts a $d$-wave altermagnetic phase~\cite{Purnendu24p263402}. 
Such a system transitions to an altermagnetic state when 
$U/t > 2.5$, consistent with our findings that a higher $U/t$ ratio drives AFM ordering, and consequently, altermagnetism. 

Furthermore, we construct phase diagrams in the presence of hole ($h$) doping for two concentrations, $p = 0.2~h$ and $p = 0.4~h$ per unit cell (uc). Hole doping significantly lowers the critical threshold for the phase transition, as shown in
Fig.~\ref{PhaseBoundary}(b-c). For $p = 0.4~h$/uc, $U^*$ decreases to approximately 1.0 eV at zero strain and drops further to 0.9 eV under a tensile strain of $\eta = 1.0$\%. These results demonstrate that strain and doping act synergistically to relax the conditions necessary for the onset of AFM ordering.

\begin{figure}[b]
\centering
\includegraphics[width=0.48\textwidth]{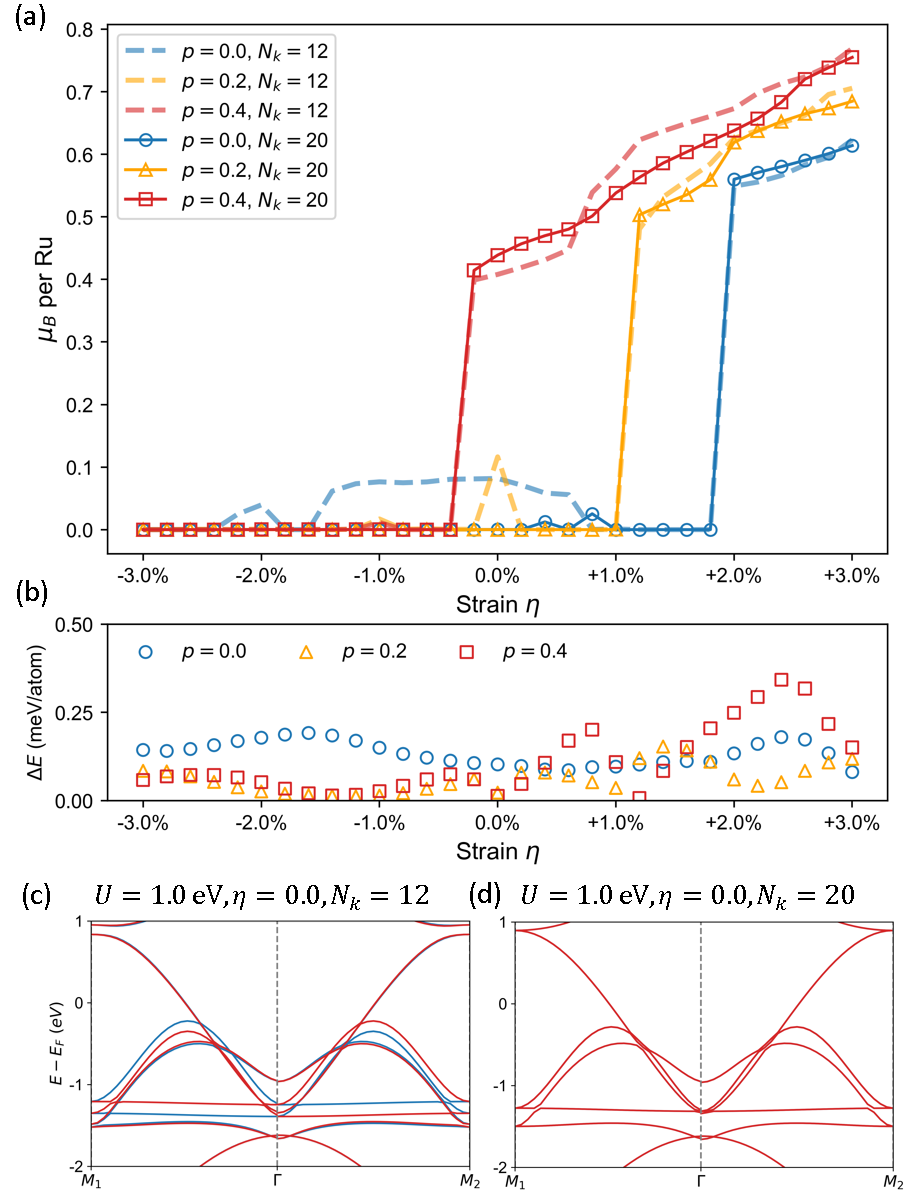}
    \caption{(a) Local magnetic moment on the Ru atom as a function of epitaxial strain and hole doping concentration, computed with two $N_k\times N_k \times N_k$ $k$-point grids and $U=1.0$~eV. (b) Energy difference ($\Delta E$) between the calculations performed with the two grids, $|E(N_k=20) - E(N_k=12)|$. Band structures of RuO$_2$ computed with (c) $N_k=12$ and (d) $N_k=20$, using $U=1.0$~eV at $\eta=0$. Spin up (down) bands are colored in red (blue). }
    \label{MEband}
\end{figure}

We also comment on the numerical fragility of the magnetic state of RuO$_2$ as determined by DFT + $U$ calculations. 
Fig.~\ref{MEband}(a) compares results computed with $U=1.0$~eV using two $N_k \times N_k \times N_k$ $k$-point grids: $N_k=12$ (solid lines with markers) and $N_k=20$ (dash lines), revealing significant discrepancies. However, the energy difference between these two $k$-point grids is less than 0.5~meV/atom [see Fig.~\ref{MEband}(b)]. This indicates that, while the total energy is well converged, the magnetic state is far more sensitive to computational parameters.
In the undoped case ($p=0.0$), the supposedly more accurate $N_k=20$ grid predicts a nonmagnetic ground state for strains $\eta < 1.8$\%, while the $N_k=12$ grid predicts a weakly AFM state with a local magnetic moment of approximately $0.1$~$\mu_B$/Ru over a broad range of $\eta$, including the zero-strain state. Since the magnitude of spin splitting scales with the local magnetic moment, the DFT + $U$ calculation with $N_k=12$ yields an altermagnetic spin-polarized band structure (see Fig.~\ref{MEband}(c)), which is absent when using the $N_k=20$ grid (see Fig.~\ref{MEband}(d)).
At a hole doping concentration of $p=0.2$~$h$/uc, both $k$-point grids predict the same critical strain value of approximately 1\% for the magnetic transition, although an anomaly is observed at the zero-strain state for $N_k=12$. 
For the higher doping concentration of $p=0.4$~$h$/uc, the critical strain value converges between the two grids. 
However, the magnitude of the local magnetic moments in the AFM state remains noticeably different.
These numerical discrepancies can be interpreted as evidence of the system's proximity to an LP instability. 
We argue that the value of $F_2^a$ depends sensitively on the DFT computational parameters, expressed as $F_2^a(U, \eta, N_k)$. 
If the true value of $F_2^a$ is near its critical threshold, numerical noise introduced by DFT calculations can significantly affect the convergence of the magnetic state. 

In summary, we explain why altermagnetism belongs to a case of unconventional magnetism, emerging as an ordered phase driven by Landau-Pomeranchuk instabilities in the spin channel.
By revisiting the historical development of unconventional magnetism within the framework of Landau-Fermi liquid theory, we aim to bridge different research communities and inspire fruitful discussions on this phenomenon. In particular, we attribute the conflicting experimental and theoretical reports on the altermagnetism of RuO$_2$ to its intrinsic proximity to a Landau-Pomeranchuk instability. This proximity makes the magnetic ground state highly sensitive to tuning parameters that influence the interaction strength among quasiparticles near the Fermi surface. Our first-principles calculations suggest that hole doping, combined with moderate tensile epitaxial strain, can effectively stabilize the AFM ordering and, consequently, the unconventional magnetism in RuO$_2$.
The susceptibility of RuO$_2$ offers a promising platform for exploring quantum phase transitions associated with Fermi surface instabilities.

{\it Acknowledgments}
C. W. is supported by the National Natural Science Foundation of China (NSFC) under the Grant No. 12234016, and also supported by the NSFC under the Grant Nos. 12174317, respectively. 
This work has been supported by the New Cornerstone Science Foundation.
The computational resource is provided by Westlake HPC center.

\bibliography{note.bib}

\section*{End Matter}

All DFT calculations are performed using the QUANTUM ESPRESSO package~\cite{Giannozzi09p395502, Giannozzi17p465901} with Optimized Norm-Conserving Vanderbilt pseudopotentials taken from the PseudoDojo library~\cite{Van18p39}. For the structural optimization of the 6-atom unit cell of rutile RuO$_2$, we employ the Perdew-Burke-Ernzerhof (PBE) exchange-correlation functional using a plane wave cutoff energy of 90 Ry, a 12$\times$12$\times$12 \textit{k}-point mesh for Brillouin zone sampling, a Gaussian smearing of 0.001 Ry, an energy convergence threshold of 10$^{-6}$ Ry, and a force convergence threshold of 10$^{-4}$ Ry/Bohr.  
Both lattice constants and ionic positions are fully relaxed. Based on the PBE optimized structure, we systematically investigate the magnetic states by applying varying magnitudes of Hubbard $U$ corrections to the Ru $4d$ states, combined with equibiaxial epitaxial strains and hole doping. We find that achieving convergence of the ground-state magnetism at certain strain states and doping conditions requires an unusually dense $k$-point grid of 20$\times$20$\times$20 and electronic self-consistency convergence threshold of 10$^{-12}$ Ry. 
The phase diagrams in Fig.~2 are constructed using these strict convergence parameters.

\end{document}